\documentstyle[11pt,newpasp,twoside,epsf]{article}
\def\edcomment#1{\iffalse\marginpar{\raggedright\sl#1\/}\else\relax\fi}
\reversemarginpar

\begin{document}
\title{The Starburst-AGN connection: Implications for the Unification of Seyferts}
 \author{Rosa M. Gonz\'alez Delgado}
\affil{Instituto de Astrof\'\i sica de Andaluc\'\i a (CSIC). Apdo. 3004, 
18080 Granada, Spain}
%\author{Ima Co-Author}
%\affil{The Name of My Institution, The Full Address of My Institution}

\begin{abstract}

Observations at ultraviolet, optical and near-infrared  wavelengths have shown 
the existence of recent star formation in the nuclear regions of Seyfert 2 (Sy2) galaxies that 
suggest a connection between the Starburst and the Seyfert phenomenon. According 
with the standard unified models of AGN circumnuclear starbursts also have to be present (and in the 
same numbers) in Sy1 as in Sy2 galaxies. This review discuss evidence in favor of the Starburst-AGN 
connection, as well as possible differences in terms of star formation activity between Sy1 and Sy2, 
that suggest an alternative interpretation of the Seyfert classification to that proposed by the 
standard unification model. Figures illustrating this contribution can be found in:
http://www.iaa.csic.es/$\sim$rosa/

\end{abstract}

\section{Introduction}

One of the most important questions concerning the active galactic nuclei (AGN) phenomenon is the 
connection between Starburst and Quasar activity. There is evidence that suggests that circumnuclear 
starbursts can have bolometric luminosities that rival even powerful Quasar. The ubiquity of 
super-massive black holes in the nuclei of normal galaxies (Kormendy \& Ho 2000) 
and the proportionality between the black hole and the spheroidal masses (Ferrarese \& Merrit 2000; 
Gerbhard et al 2000) suggest that the creation of a black hole was an integral part of 
the formation of ellipticals and the bulge of spirals. In consequence, violent events 
of star formation and AGN can coexist together and probably they did in the past even 
more often that we observe today. 

The role of starbursts in AGN has been extensively discussed theoretically in the past. 
An indirect connection between the two phenomena can exist because 
both are triggered by and live on gas fueling (mergers, interactions between galaxies, bars; e.g. Mihos \&
Hernquist 1996; Shlosman \& Noguchi, 1993). However, symbiotic processes between stars and the black hole 
(Perry \& Dyson 1985; Norman \& Scoville 1988; Willians et al 1999; Colin \& Zahn 2000), the evolution of starburst 
as the origin of the AGN activity (Terlevich \& Melnick 1985; Terlevich et al 1992), or the interaction 
of the AGN with the interstellar medium (e.g. van Breugel et al 1985; van Breugel \& Dey 1993;
Ohsuga \& Umemura 1999) are examples of a more direct connection between starbursts and AGN. 

A starburst-AGN connection is also proposed to solve a series of puzzles and 
paradoxes involving the unification model of AGN. According to the unified picture, the blue continuum 
(called the `featureless continuum', or FC) observed in Sy2 galaxies was 
believed to be scattered light from the hidden Sy1 nucleus. However, this interpretation is not completely 
correct because after the contribution of the old stars is removed, the remaining optical continuum has a 
significantly lower fractional polarization than the broad optical emission lines (Tran 1995). 
Therefore another unpolarized component has to contribute to the FC. Tran (1995) suggests that the 
unpolarized component is optically thin thermal emission from warm gas, that is heated by the central 
hidden source. In contrast, Cid-Fernandes \& Terlevich (1995) and Heckman et al. (1995) suggest that the 
unpolarized component is a young but dust-reddened stellar population, possibly associated with 
the obscuring torus. 

This paper presents evidence of the existence of young stars in the nuclei of Seyfert
and Radio galaxies that strongly suggest a connection between the two phenomena. Most of the results  
discussed here are found through the detection of stellar absorption lines produced by young stars 
in the nuclei of AGN. However, other evidence has been recently 
found  by studying the circumnuclear nebular optical emission lines in Seyferts (e.g. Goncalves \& Veron 1999; 
Kewley et al 2001).
 
\section{Evidence of young stars in low luminosity AGN: Seyfert galaxies}
   
\subsection{UV morphology}

Starbursts are very bright at ultraviolet (UV) wavelengths because they are powered by massive stars
which emit most of their flux at these and shorter wavelengths. UV images of starbursts show that
a significant fraction of the young stars form in very compact stellar clusters with size of 
a few pcs. Therefore, the UV morphology provides direct evidence of the location of the most recent 
unobscured star forming regions.

HST images of Sy2 galaxies have revealed that most of the UV continuum is produced by a 
spatially resolved nuclear starburst (Figure 1), showing subarcsec structures that may be produced by super stellar clusters.
The effective radius of the continuum emission ($\sim$100 pc) is similar to the size of the NLR 
(Gonz\'alez Delgado et al 1998).
In contract, the UV continuum light provided by the active nucleus is very little 
($\leq 20\%$, Colina et al 1997; Gonz\'alez Delgado et al 1998). 
Circumnuclear starbursts have been found also in Sy1 galaxies (e.g. NGC 7469, proposal 6358, P.I. Colina).
However, a quick analysis of HST archival data of available UV imaging of nearby Sy1, 
indicates that in many Sy1 galaxies, the UV light is emitted only by a central unresolved knot (e.g NGC 3227, 
proposal 6837, P.I. Ho) (Figure 1). This may be interpreted as a larger frequency of nuclear star formation in Sy2 
than in Sy1 and it would imply an intrinsic difference between type 1 and type 2. Therefore, it could 
present an 
evidence against the unification model. However, a systematic study of Sy1 as in Sy2 has not been done yet. 
On the other hand, a compact UV morphology by itself is not enough proof to exclude the presence of massive 
stars in the nucleus of Sy1 galaxies.

\begin{figure}
\plotone{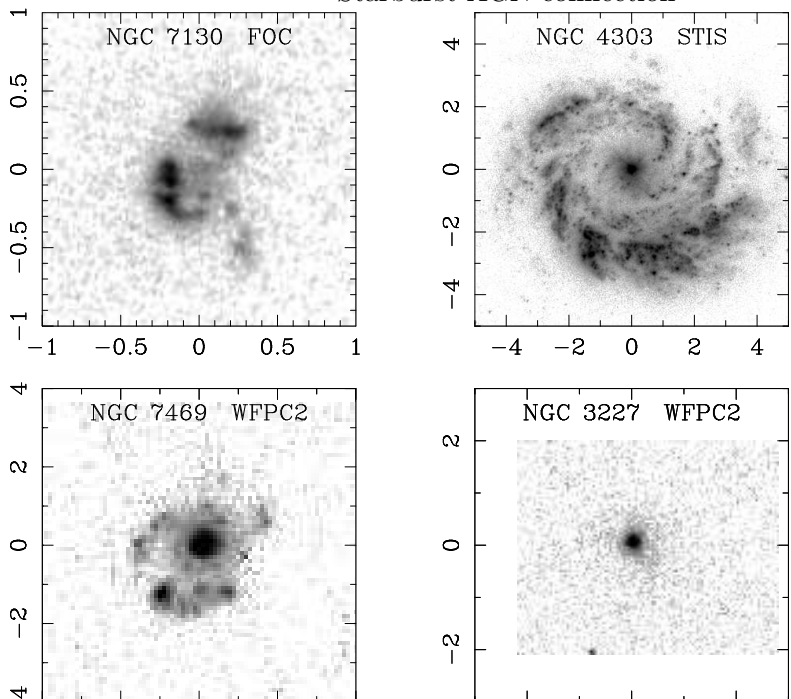}
\vspace{-6cm}
\caption{HST UV images of the Sy 2  and Sy 1 galaxies  NGC 7130, NGC 4303 and NGC 7469 and NGC 3227. 1 arcsec corresponds to 310, 80, 315 and 75 pc, respectively.} 

%\caption{HST UV images of the Seyfert 2 (NGC 7130 and NGC 4303, observed by Gonz\'alez Delgado et al 1998 and
%Colina et al 2001a, respectively) and Seyfert 1 galaxies (NGC 7469 and NGC 3227 from HST archive proposals 
%numbers 6358 and 6837, respectively). 1 arcsec corresponds to 310, 80, 315 and 75 pc, respectively.} 

\end{figure}

\subsection{UV wind lines}

A more conclusive proof that the UV continuum in AGN is provided by a starburst is obtained
by the detection of resonance wind lines (as NV $\lambda$1240, SiIV $\lambda$1400 and CIV $\lambda$1550) 
in their nuclear spectra. In starbursts, these lines show a P-Cygni profile and/or are shifted
by $\sim$ 1000-3000 km s$^{-1}$. The shape of these lines depends on the content in massive stars and  
can be used to constrain the properties (age, and the slope and upper  mass limit of the IMF) of 
the young stellar clusters. The detection of these lines in UV bright Sy2 nuclei (Figure 2  
Heckman et al 1997; Gonz\'alez Delgado et al 1998) strongly indicates that nuclear starbursts dominate 
the UV light and that they are responsible for the FC continuum in Sy2. These starbursts contribute significantly 
to the energetic of the AGN because they provide a bolometric luminosity ($\sim$ 10$^{10}$-10$^{11}$ L$\odot$) 
which is similar to the luminosity of the hidden Sy1 nucleus.
A beautiful example has been observed 
recently by Colina et al (2001a) in the Sy2 galaxy NGC 4303.
The UV image shows a circumnuclear spiraling ring of stellar clusters plus a central knot at the position
where the NICMOS images suggest the location of AGN. STIS UV spectrum of the central knot
shows strong P-Cygni wind lines that are very similar to those detected in the stellar clusters of the ring 
.
 The estimated mass (2-3 10$^{5}$ M$\odot$) and bolometric luminosity ($\sim$ 2 10$^{8}$ L$\odot$) 
of this central knot is also similar to that estimated for super stellar clusters.

\begin{figure}
\plotone{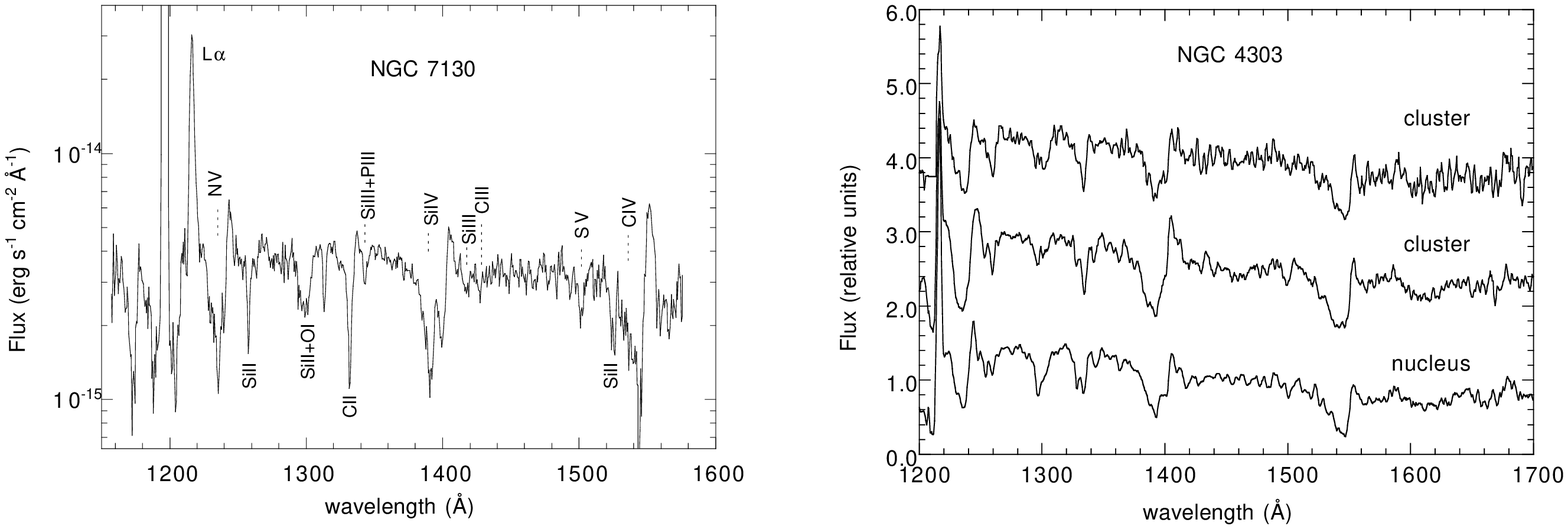}

\caption{GHRS UV nuclear (1.7$\times$1.7 arcsec) spectrum of the Sy 2 galaxy NGC 7130 (a) and  STIS UV nuclear ($\sim$20 pc) spectrum of the Sy 2 galaxy NGC 4303 and two stellar clusters (b).}

%\caption{a) GHRS UV nuclear (1.7$\times$1.7 arcsec) spectrum of the Seyfert 2 galaxy NGC 7130 observed by 
%Gonz\'alez Delgado et al (1998). b) STIS UV nuclear ($\sim$20 pc) spectrum of the Seyfert 2 galaxy 
%NGC 4303 and two stellar clusters in the ring observed by Colina et al (2001a).} 

\end{figure}

\subsection{Wolf-Rayet features}

Starbursts go through the Wolf-Rayet phase when stars more massive than 40 M$\odot$ evolve from the main
 sequence a few Myr after the onset of the burst. In this phase, the optical spectra of starbursts
show broad emission features at $\sim$ 4660 \AA, mainly due to NIII $\lambda$4634-4642
and HeII $\lambda$4686 produced by WN stars, and at 5808 \AA\ due to CIV  (if WC stars 
form in the starburst).

These features have been detected in some Sy2 (Figure 3, Heckman et al 1997; Storchi-Bergmann et al 1997;
Kunth \& Contini 1999; Tran et al 1999; Gonz\'alez Delgado et al 2001). However, the most spectacular
case has been found in Mrk 477 (Heckman et al 1997) which is the most powerful Sy2 nucleus in the local
universe. The non detection
of a broad HeII line in the polarized spectrum (Tran 1995) and the finding of the wind 
lines NV $\lambda$1240 and SiIV $\lambda$1400 suggest that the broad 4660 \AA\ feature is provided by
an ensemble of about 30000 Wolf-Rayet of a stellar cluster that is 6 Myr old.

\subsection{Near-UV and optical stellar absorption lines}

Massive stars also show photospheric absorption features (most notably the 
H Balmer series and HeI lines). Even though many of the photospheric lines in 
starbursts could be masked by the nebular lines in emission, the high-order Balmer
series (HOBS) and some of the HeI lines ($\lambda$3819, 4026, 4387 and 4922) could be 
detected in absorption. This is because the strength of the Balmer series in emission decreases rapidly
with decreasing  wavelength, whereas the equivalent width of the stellar absorption
lines is constant or increases with wavelength (Gonz\'alez Delgado, Leitherer \&
Heckman 1999). In fact, in the Sy2 in which UV stellar wind resonance lines have been detected, 
have also the HOBS and HeI in absorption.

Gonz\'alez Delgado et al (2001, GD01) and Storchi-Bergmann et al (2000, S00) have observed the near-UV spectra of 
20 Sy2 galaxies each. The targets of the two samples were selected by their nuclear nebular emission 
[OIII] $\lambda$5007 and they represent the more powerful AGN in terms of their high excitation gas.  
The criteria used to select the objects guarantee that the two samples are unbiased towards presence or absence
of starbursts. The nuclear spectra (corresponding to a few hundred pcs) show the HOBS and HeI lines in absorption 
in about 50$\%$ of the GD01 and 30$\%$ of the S00 samples (Figure 3).
Then, young and intermediate stars dominate the optical continuum in about half of these powerful Sy2 nuclei.
Recently, Cid Fernandes et al (2001; CF01) have re-analyzed the two samples using as starburst diagnostics
the equivalent width of the metallic line Ca K and two near-UV colors 
(F$_{3660}$/F$_{4020}$ and  F$_{4510}$/F$_{4020}$). They find that the nuclei that harbor a 
starburst (as detected by the UV wind lines, HOBS or Wolf-Rayet features) also have the CaK line very diluted
(Ew(Ca K) $\leq$ 10 \AA),  the old stellar population contributes less than 75$\%$ of
the optical continuum, and more than 15$\%$ of the near-UV light is provided by the starburst. In addition, in 
$\sim$ 30$\%$ of the objects, a blue continuum is required to explain the dilution of the metallic lines. 
However, in these cases the FC continuum may be due to a young stellar population that is masked by the 
old population, or is scattered light provided by the hidden Sy1 nucleus.   
These results clearly indicate that at least 40$\%$ the most powerful Sy2 harbor a nuclear starburst
and that it is responsible of the FC observed in most of these objects. On the other hand, the Sy2 with
starbursts are also the brightest galaxies at the Far-Infrared (L$_{IR} \geq 2\times 10^{10}$ L$\odot$) and
they have the larger luminosity at mid-FIR (as estimated by the 12 $\mu$m IRAS band, GD01; CF01). 
This implies that young and intermediate populations are responsible of the FIR radiation, and 
the most powerful starbursts may be associated to the most powerful AGN.

\begin{figure}
\plotone{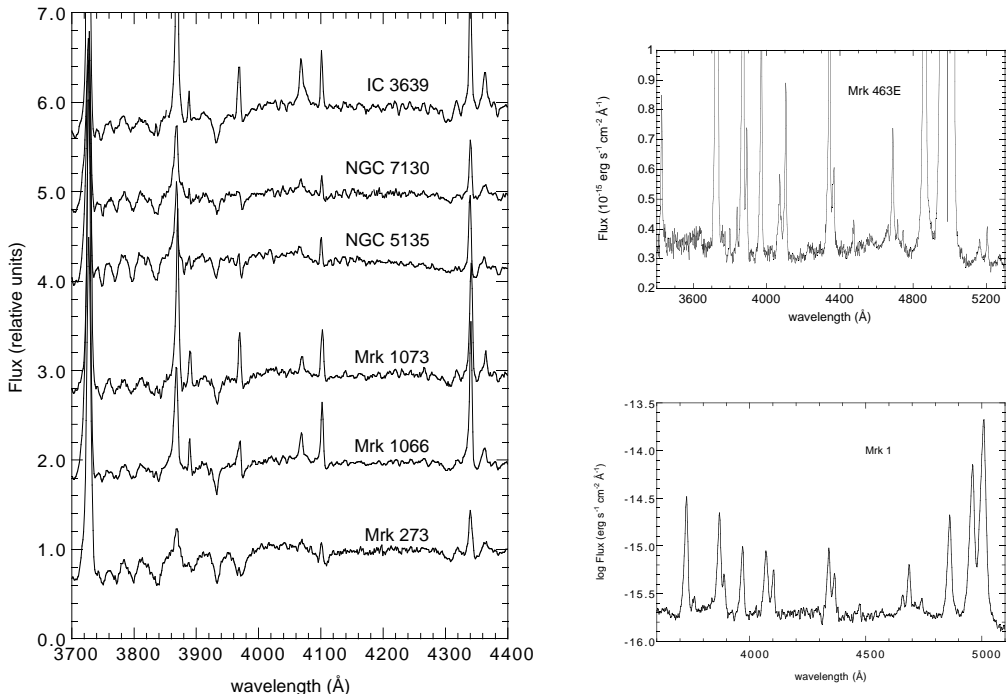}

\caption{Near-UV nuclear spectra of Sy2 galaxies that show the HOBS in absorption (a) and Wolf-Rayet features (b)}

%\caption{a) Near-UV nuclear spectra of a some Sy2 galaxies that show the HOBS and HeI lines 
%in absorption, observed by GD01 and 
%b) 4680 \AA\ broad emission feature attributed to Wolf-Rayet stars in Sy2 galaxies 
%Mrk 463E and Mrk1.  }
 
\end{figure}

%\begin{figure}
%\plotone{gonzalez-f4.eps}
%\caption{Near-UV spectrum of the QSO UN J1025-0040 observed by Brotherton et al (1999)} 
%\end{figure}

\subsection{CaII triplet}

In most of the Sy1 galaxies, the stellar features at the UV and optical wavelengths, that are very good 
diagnostics to detect young stars in Sy2 galaxies, are masked by 
the broad nebular emission lines and by the AGN continuum. However, at the near-infrared, the CaII triplet 
(CaT: $\lambda$8498, 8542 and 8662 \AA) is found in the nuclei of many Sy1 as strong as in Sy2 and normal 
galaxies. Therefore, this is a very good diagnostic to look for young stars in the nuclei of Seyfert galaxies 
and even more in Sy1. In fact, the CaT is a very good stellar diagnostic because it strongly depends 
on stellar surface gravity and to a less extent on metallicity (D\'\i az, Terlevich \& Terlevich 1989). Thus, 
in starbursts younger than 1 Gyr, the strength of the lines is very sensitive to the relative number of 
red supergiant stars (RSG) (Mayya 1997; Garc\'\i a-Vargas, Moll\'a \& Bressan 1998).

Since the pioneering work by Terlevich, D\'\i az \& Terlevich (1990), the CaT has been observed in Seyfert galaxies 
extensively. The most recent works are those of Boisson et al (2000), Jim\'enez-Benito et al (2001) and
Nelson \& Whittle (2001). In the last two works, the analysis of the CaT is combined with the Mg index (Mgb
or Mg$_2$ at $\sim$ 5200 \AA) that is a good metallicity indicator. They  
find that the CaT is very strong and the Mg lines are weak. 
For most of the Seyferts observed, the dilution by a power-law continuum can not
simultaneously explain the strength of the CaT and the weakness of the Mg index. However, starbursts younger 
than 100 Myr can account for them because the stellar population is rich in RGSs that produce a strong CaT,
and contribute to dilute the Mg lines that is produced by the
 old bulge stellar population. They conclude
that, if not all, most of the Seyfert galaxies, including 
Sy1, are composite objects consistent of an AGN plus
a starburst.

\subsection{Near-Infrared (NIR) diagnostics}

In starbursts, the NIR light is dominated by the contribution of RSG and AGB stars. The CO index (at 1.62
and 2.29 $\mu$m) is also a good young stellar diagnostic because it depends on the spectral class,
being strongest in supergiant stars (Kleinmann \& Hall 1986). However, the strength of the index also 
depends on metallicity. However, as with the CaT, at the metallicity of bulges of early type 
galaxies that are the host of Seyfert nuclei, the surface gravity dependence dominates. 
There is, however, an additional concert with this stellar diagnostic which is that the band can 
be diluted by the radiation from the warm dust that surrounds the central AGN source. 
Even so, the CO index has been successfully used to detect 
starbursts in Sy1 galaxies, where the UV light method has failed.
See for example the recent work by  Schinnerer, Eckart \& Tacconi (2001), that  
find a starburst in the inner 300 pc of the Sy1 NGC 3227 that contributes with 40$\%$ of the 
total continuum emission in the K band, although the UV image does not revealed it. 
 
However, recent works have obtained
contradictory results with respect to the estimation of the frequency of starbursts in Seyfert nuclei through 
the analysis of the NIR light. Ivanov et al (2000)  find no evidence for strong contribution of starbursts in 
the nuclei of 16 Seyfert galaxies. In contrast, Oliva et al (1999) use the stellar
index Si $\lambda$1.59 $\mu$m in addition to the CO $\lambda$1.62 and 2.29 $\mu$m to estimate the 
contribution of the non-stellar light and to measure the stellar mass to light ratio at the H band
(M/L$_H$). They find that 40$\%$ (5 of 13) of the Sy2s studied have M/L$_H$ ratio similar to starburst 
galaxies, but Sy1s (8 objects studied) have all of them a  M/L$_H$ ratio similar to elliptical galaxies. 
Thus, they conclude that old and powerful starbursts are common in Sy2 but not in Sy1. This difference
in terms of the young stellar population between Sy1 and Sy2 is in frontal conflict with the unification model
of AGN. However, the conclusion is based on a very small sample of AGN. A study based on a larger sample is under way
an it will help to check whether, as it has been proposed by Oliva and collaborators, starbursts and Seyferts
are linked by an evolutionary sequence from starburst to Sy2 to Sy1. 

\subsection{Radio Supernovas}

Another useful tool to detect starbursts is the monitoring of Seyferts to search for supernovae. 
Explosions and evolution of compact supernova remnants in the nuclei of AGN have been proposed 
by Terlevich et al (1992) to explain the origin of the BLR. In fact, Aretxaga et al (1999) have explained 
the mutation of the classical Sy2 NGC 7582 into a Sy1 by the explosion driven by a compact supernova in its
nucleus. They have also proposed this scheme to explain the variability and mutation of more than 12 
relatively low luminosity AGN.

In particular, the monitoring at radio frequencies is a powerful tool to detect starbursts in very dusty AGN. 
This technique has the advantage of not suffering from obscuretion, and sets strong
constraints on the properties of star formation in dusty systems. Colina et al (2001b) are monitoring 
several Seyferts galaxies with the VLA since 1998. They have reported the first
detection of a radio supernova in the circumnuclear region of a Seyfert galaxy. In the year 2000,
they detected at 8.4 GHz a very compact source (RSN J230315+0852.4) at 600 pc from the nucleus of the Sy1 
galaxy NGC 7469. This supernova emits 7$\%$ of the total radio flux density within 1 Kpc of the AGN.
Luminosity considerations indicate a supernova rate of 0.4 yr$^{-1}$. Even if the finding of Colina et al 
is not at all a proof for the starburst model of Terlevich and collaborator, the result indicates
that the monitoring of Seyferts at radio is a potential tool to detect starbursts very close to AGN.

\section{Evidence of young stars in high luminosity AGNs: Radio Galaxies and QSOs}

Last section has shown that nuclear starbursts are present at least in a 
significant fraction of the low luminosity regime of AGNs. Here, it is shown that young stars are also associated
to lobe-dominated radio sources and very high luminosity QSOs.

\subsection{Radio Galaxies}

FR II radio galaxies are the most luminous class of radio galaxies which host the most powerful and
extended collimated radio-jets. Many of these galaxies show a spatially extended UV morphology that is
aligned with the major axis of their radio emission. The origin of this emission has been attributed to:
1) dust and electron scattered light from an anisotropic radiation produced by the hidden AGN (in the spirit
of the unification model of AGN, Tadhunter 1987), and 2) recent star 
formation triggered by the radio source (De Young 1981). Even, there are enough proofs in favor 
of the unification model hypothesis, there are also many cases where the excess of blue color can only
be explained by young stars (e.g. Tadhunter, Dickson \& Shaw 1996). A strong evidence is shown in 
the high-z radio galaxy 4C 41.17 (Dey et al 1997). It has been found that UV light that is spatially 
aligned with the radio axis is unpolarized, and its spectrum shows wind P-Cygni and photospheric UV 
lines that are produced only by massive stars. A powerful starburst with a star formation rate 
of 140-1100 M$\odot$ yr$^{-1}$ can account for the UV continuum luminosity.
 
Aretxaga et al (2001) have observed young stars in powerful nearby radio galaxies.
They use the HOBS as a stellar diagnostic, and they detect these lines in absorption 
in 6 of the 7 galaxies observed. Their strength is much larger than that 
found in normal galaxies and they indicate the presence of a stellar population younger than 1 Gyr. 
A starburst and post-starburst stellar population have been also detected in radio (loud and quiet)
galaxies selected with very strong far-infrared emission (Tran et al 1999).

\subsection{QSOs and QSO's hosts}

Now, there is at least circumstantional evidence that suggests that there is a connection between 
starbursts and QSO activity. In fact, external processes such as galaxy interactions or mergers, have to be
invoked to sustain the high rate of gas accretion required by luminous AGNs during the life time of 
the QSO activity ($\sim$ 10$^8$ yr). Star formation could be triggered in the QSO's host and just at the
nucleus when the gas is transported and compressed to the center to feed the AGN. Ultra-luminous
far infrared galaxies (ULIRGs) may provide a clear link between galaxy mergers, starbursts and QSOs.
In these objects AGN and starburst activity co-exists together, and the dominance of one of these
two types of activity depends on the advance state of the merger (Veilleux et al 2001). Starbursts
have been found in the host of many nearby QSOs (Canalizo \& Stockton 2001). They have been 
revealed by the detection of Balmer absorption lines in the surrounding of the AGN or nebular HII emission.
The ages estimated derived for the starburst population range from a few Myr old (current star formation)
to post-starburst ages ($\leq$ 300 Myr). All these galaxies show clear signs of interactions and major
mergers, that indicate a connection between interactions, starbursts and QSO activity. Probably, the
most spectacular example is in the QSO UN J1025-0040 observed by Brotherton et al (1999). The optical
spectrum displays the broad Mg II $\lambda$2800 emission line and a strong blue continuum characteristic 
of a QSO and at the Balmer jump the higher order Balmer lines in absorption. A stellar component of about
400 Myr old with mass of 10$^{11}$ M$\odot$ is responsible for 70$\%$ of the unresolved nuclear light.
This result proves that powerful QSOs can co-exist with also very powerful starbursts just at the center
of galaxies.

\section{Impact of the starbursts on the energetics of Seyferts}

Previous sections have shown that there is strong evidence of the co-existence of starbursts and AGN in the center of
galaxies. However, the key point is to know whether the starbursts make an impact on the energetics of the AGN
or if they can even dominate over the accretion processes. Here, I concentrate to answer these points in Sy2
galaxies. A systematic study on Sy1s, similar to the one done in Sy2s, is under way and we expect to be soon 
in conditions to check whether the same conclusions are sustained for more luminous AGN.

We (GD01; CF01) have found that the most powerful Sy2 galaxies
that harbor a starburst have lower excitation (measured by [OIII] $\lambda$5007/H$\beta$), cooler
far-infrared colors (f$_{25}$/f$_{60}$) and larger mid (measured by the IRAS band at 12 $\mu$m, L$_{12}$) 
and far-infrared luminosities than the Sy2s that show an old nuclear stellar population. These differences
are expected if a starburst makes a substantial contribution to the heating of the dust and to the ionization
of the gas in the former set of nuclei, but not the latter. We found that all the Sy2s studied that have 
L$_{FIR}$ above 2$\times10^{10}$ L$\odot$ also harbor a starburst. Then, luminous far-infrared Sy2s
may owe much of their luminosity to the starburst. Another conclusion is that the Sy2s with most powerful 
nuclei may be also associated with most powerful starbursts, because the 12 $\mu$m luminosities (which arguably 
measures the hot gas heated by the AGN, Spinoglio \& Malkan 1989) are also larger in the Sy2s with young 
and intermediate age population. This luminosity link between the AGN and the starburst activity may be 
related to a common causal effect, that could be the presence of companions. In fact, 
Storchi-Bergmann et al (2001) have found that 60$\%$ of the Sy2 with nuclear starbursts have a close
companion or are in mergers, unlike to the 20$\%$ of the Sy2 nuclei with old stellar population.

Another key point is the impact of the starburst on the ionization of the gas. CF01 
have found a strong correlation between the nuclear nebular H$\beta$ luminosity in Sy2 galaxies and the 
optical continuum provided by the young stellar population or/and the FC continuum. This correlation
is similar to the Yee (1980) and Shuder (1981) correlation  that extends from Sy2s to QSOs. This correlation 
for Sy2s, that is unexpected within the prediction of the unification model of AGN, may be driven by presence 
of nuclear starbursts in Sy2s rather than by the accretion process. It also indicates that the starburst has 
a substantial impact on the ionization of the
gas. In fact, the H$\beta$ and [OIII] nebular lines in the Sy2s with nuclear starbursts are resolved into two 
components. The broad component has excitation much larger than the narrow one and it is similar to the 
excitation measured in the Sy2 nuclei with old stellar population. 
If the narrow component, that has excitation similar to typical starbursts, is produced only by 
massive stars, thus, the H$\beta$ fluxes 
indicate that the starburst contributes between 30$\%$ and 80$\%$ of
 the total 
ionization. Thus, the starbursts have a significant impact on the
emission line ratio of these Seyfert 2 nuclei. 

Levenson, Weaver \& Heckman (2001) have analyzed the X-ray spectra and images of the Sy2 galaxies that 
contain starbursts to estimate the impact of the recent star formation in X-ray. They find that although
all Seyferts of the sample show AGN at X-ray, the accretion process itself can not account for the total
X-ray emission. The emission is extended on spatial scales of $\sim$10 Kpc, which is produced by outflows
driven by the starbursts. On the other hand, the hard X-ray nuclear spectra are heavily absorbed, 
indicating that the starbursts contribute also to the obscuretion of the AGN.

\section{Implications for the Unification Models of AGNs}

Our works (SB00; GD01 and CF01) in Sy2
galaxies indicate that most of the most powerful Sy2 nuclei can be grouped into two classes, those with starbursts
and those with an old population. This segregation poses important questions: are there two different 
classes of Seyfert 2?, do all Seyfert 2 contain a hidden Seyfert 1 nucleus? In fact, the infrared colors and
excitation of the two groups are different. Nuclei with old population starbursts have [OIII]/H$\beta$ ($\sim$ 10) similar to
Sy1, and f$_{25}$/f$_{60}$ ($\sim$0.4) larger than the color suggested by Heisler, Lumsden \& Bailey (1997) for
Sy2 with no detected hidden broad emission lines (HBLR) in polarized light. In contrast, Sy2 with starbursts
have lower excitation and cooler far-infrared colors. These differences suggest that there may be two classes of Sy2:
Sy2s that have a hidden Sy1 nuclei and real Sy2 with non-HBLR that may evolve from starbursts. In this respect,
Gu et al (2001) have compiled all the available data in the literature about Seyfert galaxies with circumnuclear
starbursts (in spatial scales from few 100 pc to several kpc). They find that the Sy2 with HBLR have properties
similar to Sy1 and those non-HBLR have properties similar to starbursts. To ckeck this result, I have compiled
the spectropolarimetry observations of the samples of SB00 and 
GD01; 16 over 35 of the two samples have been observed, HBLRs have been detected in 9
of these objects. Of these 9, 5 have nuclear starbursts and 4 have old population. On the 7 objects with
non-HBLR detection, 3 have starbursts and 4 old population. This comparison suggests that there is no strong 
anti-correlation between the presence of a hidden Sy1 nucleus and a nuclear starburst. This is at odds with
a model in which one class of Sy2 nuclei are powered by hidden Sy1 nuclei and another class is powered by a 
nuclear starburst. However, Tran (2001) has suggested the existence of two intrinsically different 
populations of Sy2 galaxies: one harboring a hidden Sy1 nucleus and the other pure Sy2 without a Sy1 in which
the energetics may be dominated by a starburst component. Our previous results indicate that if in fact there 
are two different classes of Sy2 nuclei, a starburst component may be present in both classes. However, 
the starburst may also contribute to obscure the hidden Sy1. An extreme example is found in the galaxy NGC 6221
(Levenson et al 2001). Its nuclear optical spectrum is like a starburst, but its X-ray emission is like a
Sy1 showing continuum variability, a broad FeII K line and a power law continuum.

Are there differences in terms of star formation activity between Sy1 and Sy2? Gu et al (2001) based on their
literature compilation find that there is a larger fraction of star formation in Sy2 than in Sy1. On the other hand, 
based on a small statistical sample, Oliva et al (1999) arrive at the same conclusion through the estimation on
the M/L$_H$ ratio. The 'standard' unification model of AGN can not account for the apparent enhanced star
formation activity in Sy2 with respect to Sy1. As an alternative, they proposed (it was also proposed a long ago
by Heckman et al 1989) an evolutionary scheme in which Sy2 evolve into Sy1. In this scheme, starbursts obscure
the AGN, and when the massive stars explode as supernova destroy the dust that enshrouds the AGN, exposing a 
bright Sy1 nucleus. In this sense, Heckman et al (1989) proposed the Sy2 galaxies as the missing link between 
ULIRGs and QSOs. However, I think that there is not yet a strong evidence that the fraction of {\it nuclear} 
starbursts is larger in Sy2 than in Sy1, in particular considering that it is much difficult to detect nuclear
starbursts in Sy1 than in Sy2 due to the blinding glare of the AGN in Sy1 that masks most of the starburst
diagnostics used in Sy2 nuclei. However, an evolutionary scenario in which the starburst and the AGN are 
relatively long-lived, and the Sy2 with starburst can evolve to Sy2 may be an additional ingredient to the
pure geometry consideration of the standard unification model. Nuclei like Mrk 78, that has a pronounced 
post-starburst population, may represent the intermediate state. This picture suggests that nuclear starbursts
may be an integral part of the AGN phenomenon, and they are triggered at the outskirt of the molecular torus where 
gas at a few hundred parsecs of the nucleus may be accumulated and compressed to feed the AGN.

{\bf Acknowledgements} I thank to the organizers, in particular to Marco Salvati
and Roberto Maiolino for inviting me to give this talk and for their financial support. I am
also very grateful to Itziar Aretxaga and Charlie Nelson for sending me a draft of their papers 
in advance of publication, and Enrique P\'erez for his comments
from a thorough reading of the paper. I am also very grateful to my collaborators (Tim Heckman,
Thaisa Storchi-Bergmann, Roberto Cid Fernandes, Claus Leitherer, Henrique Schmitt and Luis Colina) with which 
I have done a significant fraction of the work discussed here.

\end{document}